\documentstyle[12pt,fleqn]{article}
\begin{document}

\renewcommand{\thesection}{\S\arabic{section}$\>$.}
\renewcommand{\theequation}{\arabic{equation}}
\renewcommand{\thefootnote}{\fnsymbol{footnote}\ }
\renewcommand{\thepage}{$-$\ \arabic{page}\ $-$}

\newcommand{\simlt}{\hbox{\raise3pt\hbox to 0pt{$<$}\raise-3pt\hbox{$\sim$}}}
\newcommand{\simgt}{\hbox{\raise3pt\hbox to 0pt{$>$}\raise-3pt\hbox{$\sim$}}}
\newcommand{\bfsig}{\mbox{\boldmath $\sigma$}}
\font\cmextenmagtwo=cmex10 scaled \magstep2

\baselineskip=7mm

%***************************** TITLE PAGE *************************************

$ $
\vspace*{-15mm}
\begin{flushright}
\begin{tabular}{l}
HUPD-9217 \\
December~1992
\end{tabular}
\end{flushright}

\vfil
\vfil

\begin{center}
\Large \bf

  Measuring                                 \\[3mm]
  the Beam Polarizations and the Luminosity \\[3mm]
  at Photon--Photon Colliders

\vfil
\normalsize \rm

Yoshiaki~YASUI, Isamu~WATANABE\footnotemark[1]\footnotemark[2], Jiro~KODAIRA
and Ichita~ENDO \\[7mm]
\it
Department of Physics, Hiroshima University \\[2mm]
1-3 Kagami-yama, Higashi-Hiroshima 724 JAPAN

\vfil
\vfil

\sc Abstract

\end{center}
\rm

\noindent
     We present methods to measure the beam polarizations and the luminosity
of $\gamma \gamma$ colliders at TeV energy scale.
     The beam polarizations of a $\gamma \gamma$ collider can easily be
monitored by comparing the numbers of events of the processes $\gamma \gamma$
$\rightarrow$ $\ell^+ \ell^-$ and $\gamma \gamma$ $\rightarrow$ $W^+ W^-$,
where $\ell$ means $e$ or $\mu$.
     The luminosity of a $\gamma \gamma$ collider is also measurable by the
event rate of $W$ boson pair productions and the light lepton pair
productions.

\vfil

\footnotetext[1]{
Fellow of the Japan Society for the Promotion of Science for Japanese Junior
Scientists.
Work supported in part by the Grant-in-Aid for Scientific Research from the
Ministry of Education, Science and Culture of Japan No.~040005.
}
\footnotetext[2]{
Address after Apr.~1, 1993:  Theory group, KEK, Tsukuba, Ibaraki 305, JAPAN.
}

%--------------------------------   text  -------------------------------------
\newpage
\renewcommand{\thefootnote}{\arabic{footnote})\ }
\setcounter{footnote}{0}

\section{Introduction}
\label{s1}

     Photon--photon colliders$^{[1,2]}$ are seriously considered as one of the
interesting options to upgrade future linear $e^+ e^-$ colliders.
     A lot of authors have reported on the advantages of a $\gamma \gamma$
collider from various physical viewpoints.$\!^{[3,4,5]}$
     For instance, the most interesting aspect in the $\gamma \gamma$ process
at TeV energy scale may be Higgs boson productions on the mass pole via loop
effects.
     The signal of Higgs boson productions at a photon--photon collider will
be much clearer than those at $e^+ e^-$ colliders or hadron colliders, and a
$\gamma \gamma$ collider is accessible to Higgs boson with heavier mass than
the underlying $e^+ e^-$ colliders be.$\!^{[3]}$
     As seen in this example, the photon--photon colliders will be necessary
and important facilities in the future high energy physics.

     At a photon--photon collider, the photon beams are generated through
the backward Compton scattering of the laser beams by the high energy
electron beams which are supplied by the underlying $e^+ e^-$ ($e^- e^-$)
collider.
     The theoretical studies on the beam conversion from electron into
photon$^{[1,2,6]}$ and on the photon beam collision$^{[1,2]}$ have been
performed by many authors.
     These theoretical estimates must be verified experimentally and the
luminosity and the beam polarizations should be measured by actual
observations.
     Note that both the luminosity and the polarization depend on the details
of the beam overlap at the interaction point and the knowledge of the electron
and laser polarization is not sufficient to determine the relevant parameters
of the photon-photon system.
     However the ways to measure the luminosity and the beam polarizations are
{\it not}\/ trivial at photon--photon colliders.
     Unlike the $e^+ e^-$ collisions, there is no process which has a huge
cross section, like Bhabha scattering, in $\gamma \gamma$ collisions.
     There is an idea that the luminosity could be measured by the processes
$\gamma \gamma$ $\rightarrow$ $e^+ e^- \mu^+ \mu^-$ or $e^+ e^- e^+ e^-$,
because of the constant total cross sections in the colliding
energy.$\!^{[7]}$\ \
     Unfortunately, such a claim is not realistic, since these {\it total}\/
cross sections are maintained by the collinear singularities:
     The typical muon production angle in the process $\gamma \gamma$
$\rightarrow$ $e^+ e^- \mu^+ \mu^-$ turns out to be given by the ratio of the
muon mass to the photon energy, and it is less than 10$^{-3}$~rad for the
photon energy greater than 100~GeV.
     Electrons will be produced much more collinearly.
     Such collinear leptons will never be observed by a realistic detector
which must evade the dumped bunches.

     In this paper we present a convenient technique to measure the circular
polarization of the photon beams, as well as a method of the luminosity
measurement at future photon--photon colliders.
     This paper is organized as follows.
     In {\S}2 we review the theoretical estimates on the photon beam
polarizations and the $\gamma \gamma$ luminosity.
     We derive the cross sections of lepton pair productions and $W$ boson
pair productions in {\S}3.
     The ideas of our new techniques to measure the photon beam polarization
and the collider luminosity are described in {\S}4 and {\S}5, respectively.
     {\S}6 gives some comments on the applications of our methods in the
actual experimental situations.

\section{Theories on Photon Beam Generations and Collisions}
\label{s2}

     To fix our notation, we first review the theoretical aspects of photon
beam generations and photon--photon collisions.

     The back-scattered photon spectrum can be given by the differential
Compton cross section $\sigma_c$,$\!^{[2]}$
\begin{eqnarray}
{\cal D}_{\gamma} (y)
         & \equiv &  \frac{1}{\sigma_c} \, \frac{d \sigma_c}{dy} \cr
 & & \cr
         & =      &  \frac{  \sigma'_0 + P_e P_L \sigma'_1 }%
                          {  \sigma_0  + P_e P_L \sigma_1  }  \ \ ,
\label{1}  \end{eqnarray}
where,
\begin{eqnarray}
\sigma'_0 & \equiv & \ 2 \Biggm( 1-y + \frac{1}{1-y} -4r(1-r) \Biggl) \ \ , \cr
   & & \cr
\sigma'_1 & \equiv & - 2 rx(2-y)(2r-1) \ \ , \cr
   & & \cr
\sigma_0  & \equiv & \ 2 \Biggm( 1 - \frac{4}{x} - \frac{8}{x^2} \Biggm)
                    \log (x+1) + 1 + \frac{16}{x} - \frac{1}{(x+1)^2} \ \ , \cr
   & & \cr
\sigma_1  & \equiv & \ 2 \Biggm( 1 + \frac{2}{x} \Biggm) \log (x+1)
                     - 5 + \frac{2}{x+1} - \frac{1}{(x+1)^2} \ \ .
\label{2}  \end{eqnarray}
Here $y$ is the ratio of the scattered photon energy $E_{\gamma}$ to the
electron beam energy $E_e$, {\it i.e.} $y$ $\equiv E_{\gamma}/E_e$; $x$ is the
squared ratio of the total energy of the Compton scattering in the
center-of-mass system to the electron mass $m_e$, $x$ $\equiv
4 E_e \omega/m_e^2$, where $\omega$ is the laser light energy; $r$
$\equiv y/x(1-y)$; $P_e$ and $P_L$ denote the polarizations of the beam
electron and the laser photon, respectively.
     Note here that each of the $P_e$ and $P_{\gamma}$ is normalized to $+1\
(-1)$ corresponding to the 100\% positive (negative) polarization.
     We assumed the optimum value of $x$, $x = 2 + 2\sqrt{2} \,^{[1]}$ in the
subsequent discussions.
     The value of $y$ is restricted in the range between 0 and $y_m$ $\equiv
x/(1+x)$ $\simeq 0.828$.
     The polarization distribution of the produced photon beam $P(y)$ is also
calculated.$\!^{[2]}$
\begin{equation}
P(y) \ = \
    \frac{  P_e r x \bigm[ 1 + ( 1 - y ) (  2 r -1 )^2 \bigm]
          - P_L \bigm( 1 - y +\frac{1}{1 - y} \bigm) ( 2 r -1 ) }%
         {  1 - y + \frac{1}{1 - y} - 4 r ( 1 - r )
          - P_e P_L r x ( 2 - y ) ( 2 r - 1 ) } \ \ .
\label{3}  \end{equation}
     Typical photon beam spectrum and polarization distribution are plotted in
Fig.~1 for several values of $P_e$.
     As seen in this figure, the electron beam polarization $P_e$ and the
laser photon polarization $P_L$ should be controlled such that $P_e P_L \simeq
-1$, to improve the monochromaticity of the produced photon beam.
     It is believed that the laser beam can be polarized easily and almost
completely, since the necessary laser light is visible ($E_e$ $\simlt
195$~GeV) or near infrared.
     A technology to produce a highly polarized energetic electron beam is now
developing successfully.$\!^{[8]}$
     {\it The fact that the photon beams are polarized will be an essential
feature of the $\gamma \gamma$ colliders}.
     Under such a polarization set-up,{\it i.e.} $P_e P_L$ $\simeq -1$, the
typical photon beam energy is roughly 0.8 times of the electron beam energy,
and the absolute value of $P(y)$ at large $y$ is almost uniform and unity.

     To avoid complicated discussions on the beam conversion kinematics and
collider designs, we simply evaluate the luminosity distribution as,
\begin{equation}
\frac{1}{{\cal L}_{\gamma \gamma}} \,
\frac{d^2 {\cal L}_{\gamma \gamma}}{dy_1 \, dy_2}
 \ = \  {\cal D}_{\gamma}(y_1) \; {\cal D}_{\gamma}(y_2) \ \ ,
\label{4}  \end{equation}
or,
\begin{equation}
\frac{1}{{\cal L}_{\gamma \gamma}} \,
\frac{d^2 {\cal L}_{\gamma \gamma}}{dz \, d\eta}
 \ = \  2z \, {\cal D}_{\gamma}(z {\rm e}^{+\eta}) \;
              {\cal D}_{\gamma}(z {\rm e}^{-\eta}) \ \ ,
\label{5}  \end{equation}
where $y_1$ and $y_2$ are the energy ratios of the produced photons to the
beam electrons which are accelerated to the opposite direction 1 and 2; $z$ is
the fraction of the photon--photon collision energy $\sqrt{s}$ to the sum of
the electron beam energies, {\it i.e.} $z$ $\equiv \sqrt{y_1 y_2}$
$= \sqrt{s}/2E_e$; $\eta$ is the $\gamma \gamma$ rapidity in the laboratory
system, $\eta$ $\equiv \log \sqrt{y_1/y_2}$.
     Here $z$ runs over the region between 0 and $y_m$, and then $\eta$ is
restricted to be $- \log (y_m/z)$ $\leq \ \eta$ $\leq \ + \log (y_m/z)$.
     It is assumed that the back-scattered photons collide each other just
after the Compton conversion.$\!$%
     \footnote{
     If we take into account the separation between the conversion point and
the interaction point, the effect of the finite angle of Compton scatterings
would result in an improvement of the beam monochromaticity and a suppression
of the luminosity.$\!^{[1,2]}$  }
     A contour plot of the luminosity distribution, as well as the partially
integrated luminosity distributions, is illustrated in Fig.~2 for the ideal
polarization case $P_e P_L$ $= -1$.
     A good peak is observed at the corner of high $z$ value.
     The peak in $d {\cal L}_{\gamma \gamma} / dz$ distribution is located at
$z$ $\simeq 0.788$.

     The events at low $z$ values should be discarded to avoid possible
miss-identifications of the processes.
     For example, $\gamma \gamma$ $\rightarrow$ $W^+ W^-$ $\rightarrow$
$\ell^+ \nu_{\ell} \ell^- {\bar{\nu}}_{\ell}$ or $\gamma \gamma$
$\rightarrow$ $\ell^+ \ell^-$ $+ ({\rm collinear}\, \ell^+ \ell^-)$ may be
miss-identified as low $z$ events of $\gamma \gamma$ $\rightarrow$
$\ell^+ \ell^-$, if the imbalance of transverse momenta of detected particles
is small.
     Therefore we introduce a cut on $z$ as $z$ $\geq z_{\rm cut}$ to extract
good events.
     In Table~1 and Fig.~3(a) we give the luminosity fraction ${\cal F}r$
integrated over $z_{\rm cut}$ $\leq z$ $\leq y_m$,
\begin{equation}
{\cal F}r(z_{\rm cut}) \ \equiv \
\frac{1}{{\cal L}_{\gamma \gamma}} \,
{\raise6.5mm\hbox{\cmextenmagtwo {\char'132}}}^{\ \ y_m}_{z_{\rm cut}} \!\!\!
dz \, \frac{d {\cal L}_{\gamma \gamma}}{dz} \ \ ,
\label{6}  \end{equation}
for two cases of $P_e$ $= +1.0$, $P_L$ $= -1.0$ and $P_e$ $= +0.8$,
$P_L$ $= -1.0$.
     If we adopt 0.75 as a value of $z_{\rm cut}$, ${\cal F}r(0.75)$ is 16\%
for the ideal electron beam polarization $P_e$ $= +1.0$, or is 14\% for a
conservative value of $P_e$ $= +0.8$.
     Due to the small, but finite, scattering angles of Compton
back-scatterings at the beam conversion points which would be distant from the
interaction point, the effective luminosity of a $\gamma \gamma$ collider will
be decreased from those of underlying $e^+ e^-$ colliders, though a
photon--photon collider is free from the beam--beam interaction effect which
essentially limits the luminosity of $e^+ e^-$ colliders.
     Forthcoming linear $e^+ e^-$ colliders considered are at the energy scale
between 0.3 and 2.0 TeV, and with the luminosity around 10 or 100~fb$^{-1}$ a
Snowmass year (10$^7$ s).$\!^{[9]}$\ \
     Then we focus our further discussions on future photon--photon colliders
whose center-of-mass energy is between 0.2 and 2.0 TeV and whose luminosity
above $z_{\rm cut}$ is roughly 1~fb$^{-1}$ per Snowmass year ($\rm 10^{32}
cm^{-2} s^{-1}$).

     If a beam photon is at the maximum energy fraction $y$ $= y_m$, then the
beam photon polarization $P$ is $+1$ or $-1$ according to the laser
polarization $P_L$ which is now assumed to be $\pm 1$, just as seen in
Fig.~1(b),
\begin{eqnarray}
P_L = -1 & \quad \longrightarrow \quad & P(y_m) = +1 \ , \cr
P_L = +1 & \quad \longrightarrow \quad & P(y_m) = -1 \ .
\label{7}  \end{eqnarray}
     If two photons of both beams are in this limit, the product of $\gamma$
polarizations is pure and ideal, {\it i.e.} $P_1 P_2$ $= \pm 1$, where $P_1$
and $P_2$ are the photon polarizations of beam 1 and 2, respectively.
     On the other hand, the worst combination of $P_1$ and $P_2$ above
$z_{\rm cut}$, ${\cal W}r(z_{\rm cut})$, comes from the case that one photon
has the maximum energy fraction $y$ $= y_m$ and the other has the minimum $y$
$= z_{\rm cut}^2 / y_m$,
\begin{eqnarray}
{\cal W}r(z_{\rm cut})
 & \equiv & P(y_m) \, P(z_{\rm cut}^2 / y_m) \ \ , \cr
 & =      & \pm P(z_{\rm cut}^2 / y_m) \ \ .
\label{8}  \end{eqnarray}
     The average of the polarization product $P_1 P_2$ over the whole region
of $z$ $\geq z_{\rm cut}$, ${\cal A}v(z_{\rm cut})$, can be obtained as
follows,
\begin{equation}
{\cal A}v(z_{\rm cut}) \ \equiv \
\frac{1}{{\cal L}_{\gamma \gamma}} \,
{\raise6.5mm\hbox{\cmextenmagtwo {\char'132}}}^{\ \ y_m}_{z_{\rm cut}} \!\!\!
dz \, \frac{d {\cal L}_{\gamma \gamma}}{dz} \, P_1 P_2  \ \ .
\label{9}  \end{equation}
     In Table~1 and Fig.~3(b) we also show ${\cal A}v$ and ${\cal W}r$ versus
$z_{\rm cut}$.
     If we choose $z_{\rm cut}$ $= 0.75$, the mean polarization product
${\cal A}v$ reaches to $\pm$97\% for the ideal electron polarization $P_e$
$= \pm 1.0$, and $\pm$85\% for a conservative electron polarization $P_e$
$= \pm 0.8$.
     Even in the worst polarization case ${\cal W}r(0.75)$, it drops down only
to $\pm$81\% and $\pm$59\% corresponding to $P_e$ $= \pm 1.0$ and $\pm 0.8$,
respectively.
     Therefore two colliding photon beams can be regarded as almost uniformly
polarized above the appropriate $z_{\rm cut}$ value.

\section{Lepton and $W$ Pair Productions at $\gamma \gamma$ Colliders}
\label{s3}

     To avoid the theoretical ambiguities the luminosity and the beam
polarizations will be monitored by the processes which occur at the
tree--level.
     And the processes with only two final particles will earn greater event
rates than the ones with many final particles.
     Therefore we concentrate in the processes of the light lepton pair
production $\gamma \gamma$ $\rightarrow$ $\ell^+ \ell^-$, where $\ell$ means
$e$ or $\mu$, and the W boson pair production $\gamma \gamma$ $\rightarrow$
$W^+ W^-$.
     Of course, there are some possibilities that new physics contributes to
these cross sections.
     Anomalous $\gamma WW$ or $\gamma \gamma WW$ couplings may change the $W$
pair production cross section.
     Even in the standard model, the Higgs boson pole may increase the $W$
boson production rate in the collisions of photons with the same sign
helicities.
     Such possibilities of new physics must be checked by detailed analyses.
     In the present paper, however, we confine ourselves to the standard
theory because the effects of the new physics are to be learned by analyzing
{\it the actual events} for which high statistics measurement is possible in
a reasonable period.$\!^{[4]}$

     As mentioned in the previous section the photon beams are essentially
polarized at $\gamma \gamma$ colliders.
     Then we {\it have to}\/ take into account the beam polarizations in our
calculations of the cross sections.

     The differential cross section of the process $\gamma \gamma$
$\rightarrow$ $\ell^+ \ell^-$ with the photon helicities fixed can easily be
evaluated at the tree--level.$\!^{[10]}$\ \
\begin{eqnarray}
\frac{{\rm d} \bfsig^{(\pm,\pm)}
       \mbox{\footnotesize $(\gamma \gamma \rightarrow \ell^+ \ell^-)$}}%
     {{\rm d} \cos \theta}
 & = & \frac{4 \pi \alpha^2}{s}\,
       \frac{\beta (1-\beta^4)}{(1-\beta^2 \cos^2 \theta)^2} \ \ ,\cr
 & & \cr
\frac{{\rm d} \bfsig^{(\pm,\mp)}
       \mbox{\footnotesize $(\gamma \gamma \rightarrow \ell^+ \ell^-)$}}%
     {{\rm d} \cos \theta}
 & = & \frac{4 \pi \alpha^2}{s}\,
       \frac{\beta^3 (1-\cos^2 \theta)\{2-\beta^2(1-\cos^2 \theta)\}}%
            {(1-\beta^2 \cos^2 \theta)^2} \ \ .
\label{10}  \end{eqnarray}
     Here $\beta$ and $\theta$ are the velocity and the scattering angle of
a final charged lepton~$\ell$, and $(\pm,\pm)$ or $(\pm,\mp)$ represents
that the helicities of two colliding photon are the same or the opposite
sign.
     Note here that $\beta$ and $\theta$ adopted here are defined in the
center-of-mass system of the colliding photons, not in the laboratory frame.
     The fine structure constant $\alpha$ in Eq.~(\ref{10}) should be defined
at the energy scale of $\gamma \gamma$ collisions and has a value about
$1/128$.
     For the light leptons, Eq.~(\ref{10}) is characterized by strong peaks in
the very forward and backward region caused by the collinear singularity.
     Especially in $(\pm,\pm)$ helicity set, the massless lepton can only be
emitted at $\cos \theta = \pm 1$.

     Such collinear leptons cannot be observed in an actual experimental
situation.
     Therefore, we introduce an angle cut $| \cos \theta |$ $\leq a$ and
integrate Eq.~(\ref{10}) to obtain the relevant cross section.
\begin{eqnarray}
\bfsig^{(\pm,\pm)}_{| \cos \theta | \leq a}
       \mbox{\footnotesize $(\gamma \gamma \rightarrow \ell^+ \ell^-)$}
 & = & \frac{4 \pi \alpha^2}{s}\, (1-\beta^4)
       \Biggr[ \frac{1}{2} \log \frac{1+a\beta}{1-a\beta}
            + \frac{a\beta}{1-(a\beta)^2} \Biggl] \ \ ,\cr
 & & \cr
\bfsig^{(\pm,\mp)}_{| \cos \theta | \leq a}
       \mbox{\footnotesize $(\gamma \gamma \rightarrow \ell^+ \ell^-)$}
 & = & \frac{4 \pi \alpha^2}{s}\,
       \Biggr[ \frac{5-\beta^4}{2} \log \frac{1+a\beta}{1-a\beta} \Bigg. \cr
 &   & \qquad \qquad \Bigg.
       - a\beta \biggr\{ 2 + \frac{(1-\beta^2)(3-\beta^2)}{1-(a\beta)^2}
                \biggl\} \Biggm] \ \ .
\label{11}  \end{eqnarray}
     It is possible to make a convolution of the above cross section with the
photon beam spectra.
     However, a photon beam spectrum depends on the details of the conversion
mechanics, {\it e.g.} the distance between the conversion point and the
interaction point, the Compton scattering angle, the size and the shape of the
electron bunch.
     And the beam spectrum is an object which {\it should be determined}\/ by
an actual observation experimentally.
     Therefore we do {\it not}\/ make such a convolution, and discuss in the
$\gamma \gamma$ C.~M.\ system in each collision.
     We present the cross sections with an angle cut $a = 0.9$ versus
$\sqrt{s}$ for both helicity combinations in Fig.~5, and similar ones without
any angle cut in Fig.~4 for comparison.
     One can observe a dramatic role of the angle cut in these two figures.
     As seen in Fig.~5, $e$ and $\mu$ productions with the photon helicity
$(\pm,\pm)$ are negligible for an integrated luminosity around 1~fb$^{-1}$.
     On the other hand, a large number of lepton pairs will be created by
$(\pm,\mp)$ photon beam collisions.
     It is worth noting that the production cross section of a lepton pair
are greater than those of a quark pair.
     For quark pair productions, a color factor 3 and a charge factor
(charge)$^4$ should be multiplied to R.~H.~S.$\;$of Eqs.~(\ref{10}) and
(\ref{11}), rather than $3 \times (\rm charge)^2$ in $e^+ e^-$ collisions.

     The cross section of the process $\gamma \gamma$ $\rightarrow$ $W^+ W^-$
shows a different dependence on the photon helicities.
     The differential cross sections and the cross sections with the angle cut
for both helicity combinations are as follows.$\!^{[11]}$
\begin{eqnarray}
\frac{{\rm d} \bfsig^{(\pm,\pm)}
              \mbox{\footnotesize $(\gamma \gamma \rightarrow W^+ W^-)$}}%
     {{\rm d} \cos \theta}
 & = & \frac{2 \pi \alpha^2}{s}\,
       \frac{\beta_W (3+\beta_W^2) (1+3\beta_W^2)}%
            {(1-\beta_W^2 \cos^2 \theta)^2} \ \ ,\cr
 & & \cr
\frac{{\rm d} \bfsig^{(\pm,\mp)}
              \mbox{\footnotesize $(\gamma \gamma \rightarrow W^+ W^-)$}}%
     {{\rm d} \cos \theta}
 & = & \frac{2 \pi \alpha^2}{s}\, \beta_W
       \Biggr[ 3 - \frac{22-6\beta_W^2}{1-\beta_W^2 \cos^2 \theta} \cr
 &   & \qquad \qquad \quad
                 + \frac{(5-\beta_W^2) (7-3\beta_W^2)}%
                        {(1-\beta_W^2 \cos^2 \theta)^2} \Biggl] \ \ ,
\label{12}  \end{eqnarray}
and,
\begin{eqnarray}
\bfsig^{(\pm,\pm)}_{| \cos \theta | \leq a}
       \mbox{\footnotesize $(\gamma \gamma \rightarrow W^+ W^-)$}
 & = & \frac{2 \pi \alpha^2}{s}\, (3+\beta_W^2) (1+3\beta_W^2) \cr
 &   & \qquad \Biggr[ \frac{1}{2} \log \frac{1+a\beta_W}{1-a\beta_W}
              + \frac{a\beta_W}{1-(a\beta_W)^2} \Biggl] \ \ ,\cr
 & & \cr
\bfsig^{(\pm,\mp)}_{| \cos \theta | \leq a}
       \mbox{\footnotesize $(\gamma \gamma \rightarrow W^+ W^-)$}
 & = & \frac{2 \pi \alpha^2}{s}\,
       \Biggr[ 6a\beta_W
             + (5-\beta_W^2) (7-3\beta_W^2) \frac{a\beta_W}{1-(a\beta_W)^2}
       \Bigg. \cr
 &   & \qquad \qquad \Bigg.
       - \frac{9+10\beta_W^2-3\beta_W^4}{2}
         \log \frac{1+a\beta_W}{1-a\beta_W} \Biggl] \ \ .
\label{13}  \end{eqnarray}
     Here $\beta_W$ is the velocity of the final $W$ boson.
     The center-of-mass energy dependences of the above cross sections can
also be found in Figs.~4 and 5.
     The cross section of this process shows a mild dependence on the helicity
set, even if one introduces an angle cut.

\section{Polarization Measurement of Photon Beams}
\label{s4}

     The fact that the cross sections of two processes $\gamma \gamma$
$\rightarrow$ $\ell^+ \ell^-$ and $\rightarrow$ $W^+ W^-$ show different
dependences on the photon helicities gives us an idea to measure the
polarizations of the photon beams.
     We introduce a ratio of these two production cross sections,
\begin{equation}
{\cal R}_{\ell/W} \ \equiv \
     \frac{\bfsig_{| \cos \theta | \leq a}
           \mbox{\footnotesize $(\gamma \gamma \rightarrow \ell^+ \ell^-)$}}%
          {\bfsig_{| \cos \theta | \leq a}
           \mbox{\footnotesize $(\gamma \gamma \rightarrow W^+ W^-)$}} \ \ .
\label{14}  \end{equation}
     The $\sqrt{s}$ dependences of ${\cal R}_{\ell/W}$ for both photon
helicity combinations with $a$ = 0.9 are represented in Fig.~6.
     For a set of partly polarized photon beams the cross section ratio
${\cal R}_{\ell/W}$ is, of course, described as follows,
\begin{equation}
{\cal R}_{\ell/W} \ = \
     \frac{ \frac{1 + P_1 P_2}{2}
            \bfsig_{| \cos \theta | \leq a}^{(\pm,\pm)}
            \mbox{\footnotesize $(\gamma \gamma \rightarrow \ell^+ \ell^-)$}
          + \frac{1 - P_1 P_2}{2}
            \bfsig_{| \cos \theta | \leq a}^{(\pm,\mp)}
            \mbox{\footnotesize $(\gamma \gamma \rightarrow \ell^+ \ell^-)$}}%
          { \frac{1 + P_1 P_2}{2}
            \bfsig_{| \cos \theta | \leq a}^{(\pm,\pm)}
            \mbox{\footnotesize $(\gamma \gamma \rightarrow W^+ W^-)$}
          + \frac{1 - P_1 P_2}{2}
            \bfsig_{| \cos \theta | \leq a}^{(\pm,\mp)}
            \mbox{\footnotesize $(\gamma \gamma \rightarrow W^+ W^-)$}}
\ , \
\label{15}  \end{equation}
or more simply, neglecting $\bfsig_{| \cos \theta | \leq a}^{(\pm,\pm)}
\mbox{\footnotesize $(\gamma \gamma \rightarrow \ell^+ \ell^-)$}$,
\begin{equation}
{\cal R}_{\ell/W} \ \simeq \
     \frac{ \frac{1 - P_1 P_2}{2}
            \bfsig_{| \cos \theta | \leq a}^{(\pm,\mp)}
            \mbox{\footnotesize $(\gamma \gamma \rightarrow \ell^+ \ell^-)$}}%
          { \frac{1 + P_1 P_2}{2}
            \bfsig_{| \cos \theta | \leq a}^{(\pm,\pm)}
            \mbox{\footnotesize $(\gamma \gamma \rightarrow W^+ W^-)$}
          + \frac{1 - P_1 P_2}{2}
            \bfsig_{| \cos \theta | \leq a}^{(\pm,\mp)}
            \mbox{\footnotesize $(\gamma \gamma \rightarrow W^+ W^-)$}}
\ , \
\label{16}  \end{equation}
where $P_1$, $P_2$ are each {\it average}\/ polarization of two photon beams 1
and 2 under the condition $z$ $\geq z_{\rm cut}$.
     The $P_1 P_2$ dependences of ${\cal R}_{\ell/W}$ at several values of
$\sqrt{s}$ are plotted in Fig.~7.
     The ratio ${\cal R}_{\ell/W}$ is roughly proportional to $1 - P_1 P_2$,
because the denominator in Eq.~(\ref{16}) is not so variant against a change
of the colliding photon helicity combination, as seen in Fig.~5.

     In an actual experiment ${\cal R}_{\ell/W}$ can be determined by the
event rate of two processes,
\begin{equation}
{\cal R}_{\ell/W} \ = \
 \frac{N_{| \cos \theta | \leq a}
 \mbox{\footnotesize ($\gamma \gamma \rightarrow \ell^+ \ell^-$)}}%
      {N_{| \cos \theta | \leq a}
 \mbox{\footnotesize ($\gamma \gamma \rightarrow W^+ W^- \rightarrow$ 4jets)}}
 \ Br \mbox{\footnotesize ($W \rightarrow$ 2jets)}^2 \ \ ,
\label{17}  \end{equation}
where $N_{| \cos \theta | \leq a} \mbox{\footnotesize ($\gamma \gamma
\rightarrow \ell^+ \ell^-$)}$ is the number of events of the lepton pair
production within $| \cos \theta |$ $\leq a$ and
$N_{| \cos \theta | \leq a} \mbox{\footnotesize ($\gamma \gamma \rightarrow
W^+ W^- \rightarrow$ 4jets)}$ is the number of events of four quark jets via a
$W$ pair production within $| \cos \theta |$ $\leq a$.
     And $Br\mbox{\footnotesize ($W \rightarrow$ 2jets)}$ is the branching
ratio of $W$ boson decay into two quark jets, which has the value $\sim 2/3$
in the standard model.
     Here the reason why we concentrate on the neutrinoless final states is to
determine $z$ by the energy sum of the final states.

     The cross sections for the helicity sets $(\pm,\mp)$ are large enough to
measure the polarization product $P_1 P_2$ precisely.
     For example, photon--photon collisions above $z_{\rm cut}$ in an
operation of 1~fb$^{-1}$ integrated luminosity, corresponding to 1 Snowmass
year run, at $\sqrt{s}$ =~400~GeV with 100\% $(\pm,\mp)$ photon beam
polarizations would produce about 7,600 lepton pairs for each species, and
about 14,600 $W$ pairs decaying into quark jets at $a$ = 0.9.
     As mentioned in {\S}2, the beam polarizations can be regarded as uniform
if we adopt an appropriate $z_{\rm cut}$ value.
     Then we can treat whole events altogether in an evaluation of $P_1 P_2$,
and we obtain the expected statistical error of ${\cal R}_{\ell/W}$ less than
$3 \times 10^{-3}$ for any value of $P_1 P_2$.
     The corresponding maximum statistical error of $P_1 P_2$ is only 3{\%}.

\section{Luminosity Measurement at Photon--Photon Colliders}
\label{s5}

     Due to a wide spread of the colliding $\gamma \gamma$ invariant mass as
seen in Fig.~2(b), one should determine not only the total luminosity, but
also the luminosity distribution on the $\gamma \gamma$ invariant mass.

     To monitor the luminosity distribution in an actual experiment, one can
use the process with the largest cross section with an angle cut.
     The integrated luminosity $\int {\cal L}_{\gamma \gamma} \, dt$ can be
evaluated at each $z$ value,
\begin{eqnarray}
\hspace*{-7mm}
\int {\cal L}_{\gamma \gamma} \, dt \ = & \hspace*{10cm} \nonumber
\end{eqnarray}
\begin{equation}
\hspace*{-1cm} \begin{array}{c}
 N_{| \cos \theta | \leq a}
 \mbox{\footnotesize ($\gamma \gamma \rightarrow W^+ W^- \rightarrow$ 4jets)}
\\ \hline
 \mbox{\small $Br\mbox{\footnotesize ($W \rightarrow$ 2jets)}^2$}
 \bigm[  \mbox{\small $\frac{1 + P_1 P_2}{2}$}
         \mbox{\small $\bfsig^{(\pm,\pm)}_{| \cos \theta | \leq a}$}
         \mbox{\scriptsize ($\gamma \gamma \rightarrow W^+ W^-$)}
       + \mbox{\small $\frac{1 - P_1 P_2}{2}$}
         \mbox{\small $\bfsig^{(\pm,\mp)}_{| \cos \theta | \leq a}$}
         \mbox{\scriptsize ($\gamma \gamma \rightarrow W^+ W^-$)}
  \bigl]
\end{array}  \ \ .
\label{18}  \end{equation}

     The statistical fluctuation of the number of four-jet events prevails
the statistical error of the measured integrated luminosity.
     It is 8\% for a one-day operation and 0.8\% for a year operation at
$\sqrt{s}$ = 400~GeV, where ${\cal L_{\gamma \gamma}}$ = 1~fb$^{-1}$ a year
above $z_{\rm cut}$ and the standard model branching ratio are assumed.
     A similar formula can be derived for the lepton pair productions, however
the lepton pair productions have the disadvantage in the event rate at
$\sqrt{s}$ $\simgt 300$~GeV.

\section{Discussions}
\label{s6}

     Our methods presented in the above sections are based on the tree--level
cross sections of the standard model without Higgs resonance.
     As mentioned in {\S}3, the cross sections Eqs.~(\ref{10})$\sim$(\ref{13})
may be shifted by the radiative corrections, the Higgs contribution, as well
as new physics like anomalous $\gamma WW$, $\gamma \gamma WW$ couplings, new
particle resonances or $W$ boson rescatterings by new strong forces, and so
on.
     Such possibilities must be checked by more detailed physical
considerations, including the analyses of the event topology and comparisons
with the theoretical simulations of the new physics.
     We think that improvements to incorporate these effects into our
techniques are trivial and easily performed, if the deviations from the
standard model caused by the new physics are not so large.
     The top pair production can be available without worrying about
ambiguities of the gauge self-couplings, however, the cross section is not so
large, as seen in Fig.~5.

     The technique presented in {\S}4 is {\it not}\/ measuring each photon
polarization $P_1$ and $P_2$ independently, but measuring only their product
$P_1 P_2$.
     A separate measurement of $P_1$ and $P_2$ may be performed through other
processes with more complicated final states.

     In conclusion, the photon beam polarizations and the luminosity at a
photon--photon collider can be measured by looking at of both processes
$\gamma \gamma$ $\rightarrow$ $\ell^+ \ell^-$ and $\gamma \gamma$
$\rightarrow$ $W^+ W^-$.
     It is shown that the beam polarization product and the luminosity can be
measured within a sufficiently short time with a good accuracy.
     We believe that the methods proposed in this paper will be powerful tools
to understand the photon beam features at future $\gamma \gamma$ colliders.

%---------------------------   acknowledgments   ------------------------------
\renewcommand{\thesection}{}

\section{Acknowledgments}

     We would like to acknowledge valuable discussions with
Drs.~Hirotomo~IWASAKI, Akiya~MIYAMOTO and Toshiaki~TAUCHI.

%------------------------------   references   --------------------------------
\newpage

\section{References}

\begin{description}

%1% photon-photon colliders, unpolarized
\item[{[1]}]   I.~F.~Ginzburg, G.~L.~Kotkin, V.~G.~Serbo and V.~I.~Tel'nov,
                             {\it Pis'ma Zh.\ Eksp.\ Teor.\ Fiz.}, {\bf 34}
                             (1981) 514 [{\it JETP Lett.}, {\bf 34} (1982)
                             491];
                             {\it Nucl.\ Instrum.\ Methods}, {\bf 205} (1983)
                             47.

%2% polarization control, energies and luminosities of photon-photon colliders
\item[{[2]}]   I.~F.~Ginzburg, G.~L.~Kotkin, S.~L.~Panfil, V.~G.~Serbo and
                             V.~I.~Telnov, {\it Nucl.\ Instrum.\ Methods
                             Phys.\ Res.}, {\bf 219} (1984) 5. \\
\hspace*{-4mm} V.~I.~Telnov, {\it Nucl.\ Instrum.\ Methods Phys.\ Res},
                             {\bf A294} (1990) 72. \\
\hspace*{-4mm} I.~Endo, {\it Proc.$\;$of the Second Workshop on Japan Linear
                             Collider (JLC)}, KEK, November 6--8, 1990,
                             pp.~323, ed.\ S.~Kawabata, KEK Proceedings
                             91-10 (1991). \\
\hspace*{-4mm} D.~L.~Borden, D.~A.~Bauer and D.~O.~Caldwell,
                             {\it SLAC preprint}, SLAC-PUB-5715 (1992).
%%%%%%%%%%%%%%%%%%%%%%%%%%   {\it Phys.\ Rev.} {\bf D}

%3% photon-photon physics
%%% Higgs production
\item[{[3]}]   R.~Najima, {\it Proc.$\;$of the Third Meeting on Physics at
                             Te$\!$V Energy Scale}, KEK, September 28--30,
                             1989, pp.~112, ed.\ K.~Hikasa and C.~S.~Lim, KEK
                             Report 90-9 (1990). \\
\hspace*{-4mm} T.~L.~Barklow, {\it SLAC Preprint}, SLAC-PUB-5364 (1990). \\
\hspace*{-4mm} E.~E.~Boose and G.~V.~Jikia, {\it Phys.\ Lett.}, {\bf 275}
                             (1992) 164.

%4% photon-photon physics
%%% gamma-W anomalous couplings
\item[{[4]}]   I.~F.~Ginzburg, G.~L.~Kotkin, S.~L.~Panfil and V.~G.~Serbo,
                             {\it Nucl.\ Phys.}, {\bf B228} (1983) 285. \\
\hspace*{-4mm} E.~Yehudai, {\it Phys.\ Rev.}, {\bf D44} (1991) 3434. \\
\hspace*{-4mm} S.~Y.~Choi and F.~Schrempp, {\it DESY Preprint}, DESY 91-155
                             (1991).

%5% photon-photon physics
%%% top production
\item[{[5]}]   O.~J.~P.~\'Eboli, M.~C.~Gonzalez-Garcia, F.~Halzen and
                             S.~F.~Novaes, {\it Univ.$\;$of Wisconsin --
                             Madison Preprint}, MAD/PH/701, IFT-P.014/92
                             (1992). \\
\hspace*{-4mm} J.~H.~K\"uhn, E.~Mirkes and J.~Steegborn, {\it Universit\"at
                             Karlsruhe Preprint}, TTP92-28 (1992). \\
%%% top-Higgs coupling
\hspace*{-4mm} E.~Boos, I.~Ginzburg, K.~Melnikov, T.~Sack and S.~Shichanin,
                             {\it Max-Planck-Institut f\"ur Physik Preprint},
                             MPI-Ph/92-48 (1992).  \\
%%% heavy fermion-Higgs coupling
\hspace*{-4mm} M.~S.~Chanowitz, {\it Phys.\ Rev.\ Lett.}, {\bf 69} (1992)
                             2037. \\
%%% susy physics
\hspace*{-4mm} A.~Goto and T.~Kon, {\it Europhysics Lett.}, {\bf 13} (1990)
                             211; erratum {\it ibid.}, {\bf 14} (1991) 281. \\
%%% summary
\hspace*{-4mm} D.~L.~Borden {\it et al.}\/ in Ref.~[2].

%6% beam conversion
\item[{[6]}]   R.~H.~Milburn, {\it Phys.\ Rev.\ Lett.}, {\bf 10} (1963) 75. \\
\hspace*{-4mm} F.~R.~Arutyunyan and V.~A.~Tumanyan, {\it Zh.\ Eksp.\ Theor.\
                             Fiz.}, {\bf 44} (1963) 2100.  [{\it Sov.\ Phys.\
                             JETP}, {\bf 17} (1963) 1412].

%7% luminosity measurement
\item[{[7]}]   I.~F.~Ginzburg {\it et al.}\/ in Refs.~[1] and [2].

%8% e beam polarization
\item[{[8]}]   T.~Nakanishi, H.~Aoyagi, H.~Horinaka, Y.~Kamiya, T.~Kato,
                             S.~Nakamura, T.~Saka and M.~Tsubata, {\it Phys.\
                             Lett.}, {\bf A158} (1991) 345. \\
\hspace*{-4mm} H.~Aoyagi, H.~Horinaka, Y.~Kamiya, T.~Kato, T.~Kosugoh,
                             S.~Nakamura, T.~Nakanishi, S.~Okumi, T.~Saka,
                             M.~Tawada and M.~Tsubata, {\it Phys.\
                             Lett.}, {\bf A167} (1992) 415. \\
\hspace*{-4mm} T.~Murayama, E.~L.~Garwin, R.~Prepost, G.~H.~Zapalac,
                             J.~S.~Smith and \\
                             J.~D.~Walker, {\it Phys.\ Rev.\ Lett.}, {\bf 66}
                             (1991) 2376.

%9% e+ e- collier parameters
\item[{[9]}]   {\it Proc.$\;$of the Second Workshop on Japan Linear Collider
                             (JLC)}, KEK, November 6--8, 1990, edited by
                             S.~Kawabata, KEK, Proceedings 91-10 (1991). \\
\hspace*{-4mm} S.~Orito, {\it Univ.$\;$of Tokyo, ICEPP preprint}, UT-ICEPP
                             92-05 (1992). \\
\hspace*{-4mm} {\it Proc.$\;$of the International Workshop on Next--Generation
                             Linear Colliders}, Stanford, California, edited
                             by M.~Riordan, {\it SLAC preprint}, SLAC-335
                             (1988). \\
\hspace*{-4mm} R.~Ruth, {\it SLAC preprint}, SLAC-PUB-5406 (1991). \\
\hspace*{-4mm} T.~Weiland, to appear in {\it Proc.$\;$of the 1991 Conference
                             on Physics at Linear Colliders}, Saariselka,
                             Finland. \\
\hspace*{-4mm} {\it Proc.$\;$of the First International TESLA Workshop},
                             Ithaca, New York, edited by H.~Padamsee, {\it
                             Cornell preprint}, CLNS-90-1029 (1990). \\
\hspace*{-4mm} P.~B.~Palmer, {\it Ann.\ Rev.\ Nucl.\ Part.\ Sci.}, {\bf 40}
                             (1990) 529.

%10% lepton pair production
\item[{[10]}]   An essentially equivalent formula of the differential cross
                             section of fermion pair production from the
                             polarized photons can be found in T.~L.~Barklow
                             in Ref.~[3].

%11% W pair production
\item[{[11]}]   A formula of the differential cross section on Mandelstam $t$
                             variable of $W$ pair production from the
                             polarized photons can be found in I.~F.~Ginzburg
                             {\it et al.} in Ref.~[4].

\end{description}

%---------------------------------   table   ----------------------------------
\newpage

\section{Table}

\vfil

\begin{table}[h]
\noindent
{\bf Table~1:} \quad
    The luminosity fraction ${\cal F}r$, the average ${\cal A}v$ of the
polarization product $P_1 P_2$ and the minimum ${\cal W}r$ of the polarization
product above $z_{\rm cut}$ in case of both photon beams are generated by a
electron beam with $P_e$ $= +1.0$ or $+0.8$, and a laser beam $P_L$ $= -1.0$,
at several values of $z_{\rm cut}$.  The optimum laser light energy
($x = 2+2\sqrt{2}$) is assumed.  \\[5mm]
\begin{center}
\begin{tabular}{|c||c|c|c||c|c|c|}
\hline
 & \multicolumn{3}{c||}{$P_e = +1.0$} & \multicolumn{3}{c|}{$P_e = +0.8$} \\
\hline
 $z_{\rm cut}$ & ${\cal F}r$ & ${\cal A}v$ & ${\cal W}r$
               & ${\cal F}r$ & ${\cal A}v$ & ${\cal W}r$ \\
\hline \hline
    0.80  &  0.045  &  0.997  &  0.985  &  0.036  &  0.953  &  0.913  \\
    0.79  &  0.070  &  0.991  &  0.970  &  0.057  &  0.936  &  0.868  \\
    0.78  &  0.095  &  0.987  &  0.946  &  0.078  &  0.920  &  0.813  \\
    0.77  &  0.120  &  0.983  &  0.911  &  0.099  &  0.897  &  0.747  \\
    0.76  &  0.143  &  0.974  &  0.866  &  0.119  &  0.877  &  0.672  \\
    0.75  &  0.164  &  0.965  &  0.807  &  0.138  &  0.853  &  0.586  \\
    0.74  &  0.184  &  0.955  &  0.736  &  0.156  &  0.830  &  0.492  \\
    0.73  &  0.203  &  0.943  &  0.651  &  0.172  &  0.805  &  0.391  \\
    0.72  &  0.220  &  0.928  &  0.554  &  0.188  &  0.780  &  0.285  \\
    0.71  &  0.236  &  0.912  &  0.447  &  0.203  &  0.753  &  0.176  \\
    0.70  &  0.250  &  0.896  &  0.333  &  0.218  &  0.726  &  0.067  \\
\hline
\end{tabular}
\end{center}
\end{table}

\vfil
\vfil
\vfil
\vfil

%----------------------------   figure captions   -----------------------------
\newpage

\section{Figure Captions}

\begin{description}

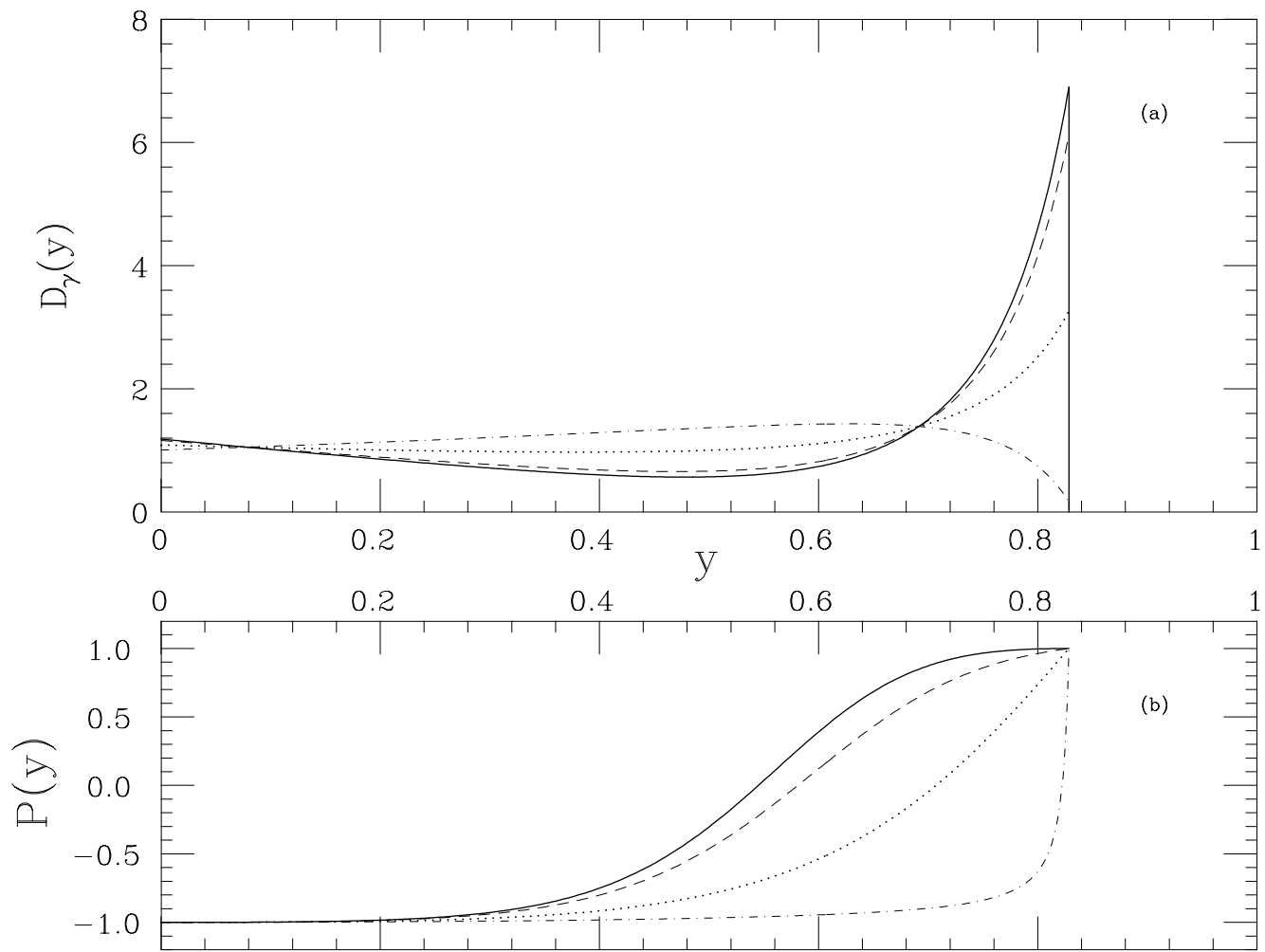
\item[Fig.~1]  The photon beam spectra (a) and the polarization distributions
     (b) for the electron beam polarization $P_e$ $= +1.0$ (solid line),
     $+0.8$ (dashed line), $0.0$ (dotted line) and $-1.0$ (dot-dashed line).
     The laser beam polarization is fixed to be $P_L$ $= -1.0$ for each $P_e$,
     and the optimum laser light energy ($x = 2+2\sqrt{2}$) is assumed.

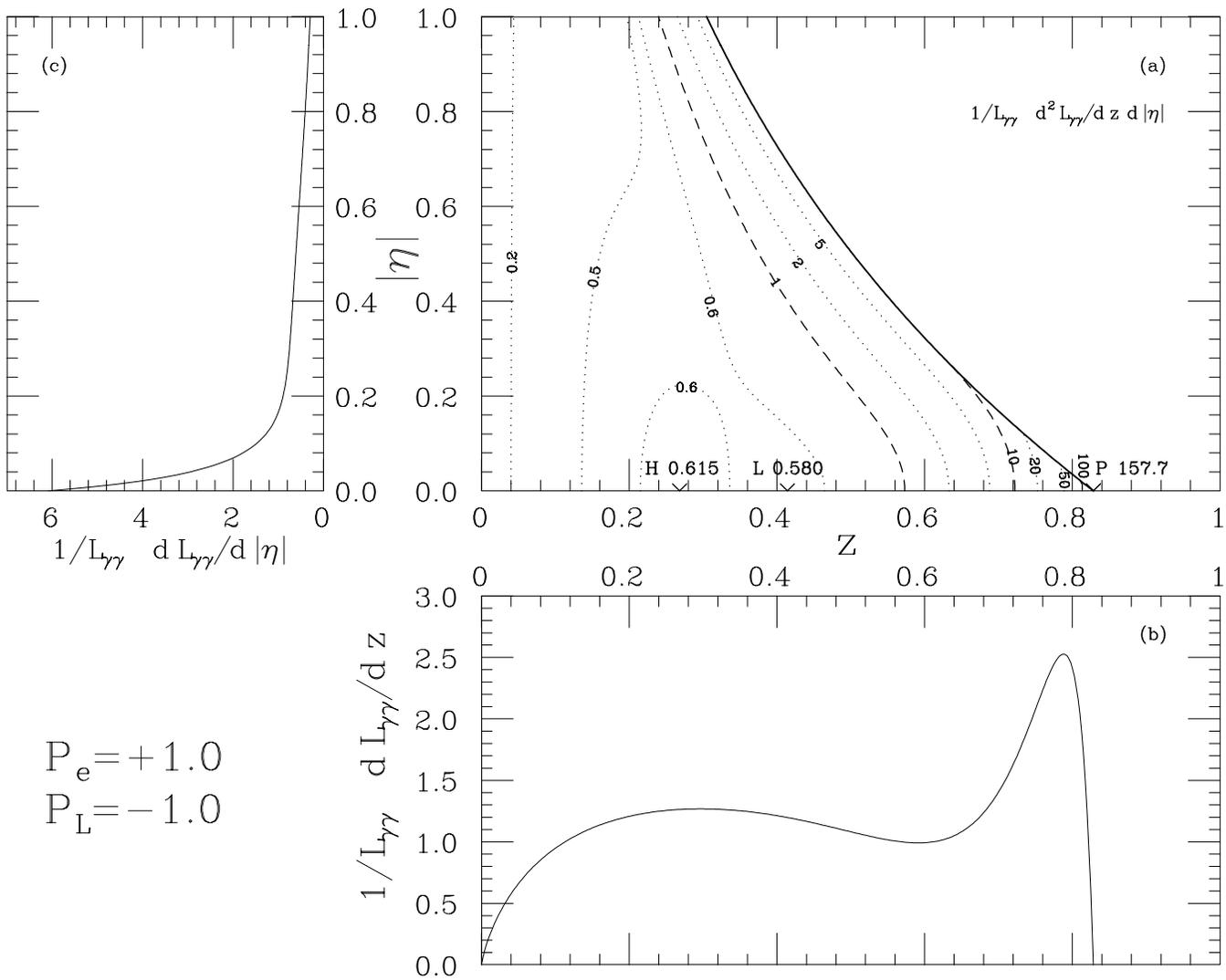
\item[Fig.~2]  A contour plot (a) of the luminosity distribution on the energy
     fraction $z$ and the {\it absolute value}\/ of $\gamma \gamma$ rapidity
     $|\eta|$ for the ideal combination of the electron beam and the laser
     polarizations $P_e P_L = -1$.  `V' marks are plotted at the peak point,
     a local maximum and a local minimum with the code P, H and L,
     respectively, as well as the value of the distribution at these
     extremes.  Values on the contour lines are 0.2, 0.5, 0.6 (two distinct
     lines), 1, 2, 5, 10, 20, 50 and 100 from left to right, respectively.
     The lines of the value 1, 10 and 100 are dashed and else are dotted.  The
     range $|\eta| > 1$ is omitted, and the area at right of the bold-solid
     line is not allowed.  Semi-integrated distributions
     $1/{\cal L}_{\gamma \gamma} \ \,d{\cal L}_{\gamma \gamma} / d z$ (b) and
     $1/{\cal L}_{\gamma \gamma} \ \,d{\cal L}_{\gamma \gamma} / d |\eta|$
     (c) are also displayed.  The optimum laser light energy ($x =
     2+2\sqrt{2}$) is assumed.

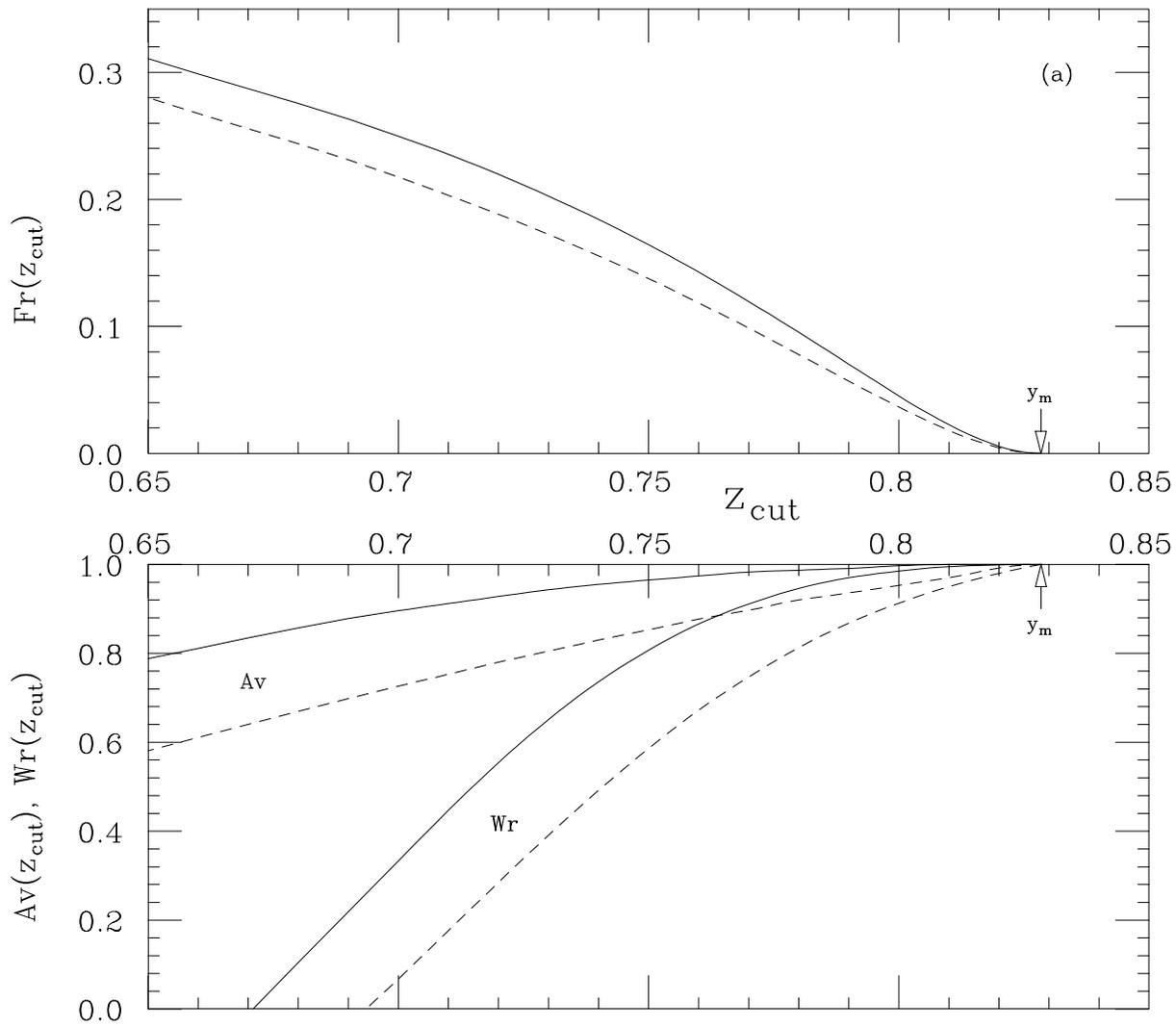
\item[Fig.~3]  The $z_{\rm cut}$ dependences of the luminosity fraction
     ${\cal F}r$ (a), the average ${\cal A}v$ of the polarization product
     $P_1 P_2$ (upper lines in (b)) and the minimum ${\cal W}r$ of the
     polarization product (lower lines in (b)) above $z_{\rm cut}$.  The solid
     lines correspond to $P_e$ $= +1.0$, while the dashed lines are for $P_e$
     $= +0.8$.  The laser beam polarization is fixed to be $P_L$ $= -1.0$ for
     each $P_e$, and the optimum laser light energy ($x = 2+2\sqrt{2}$) is
     assumed.

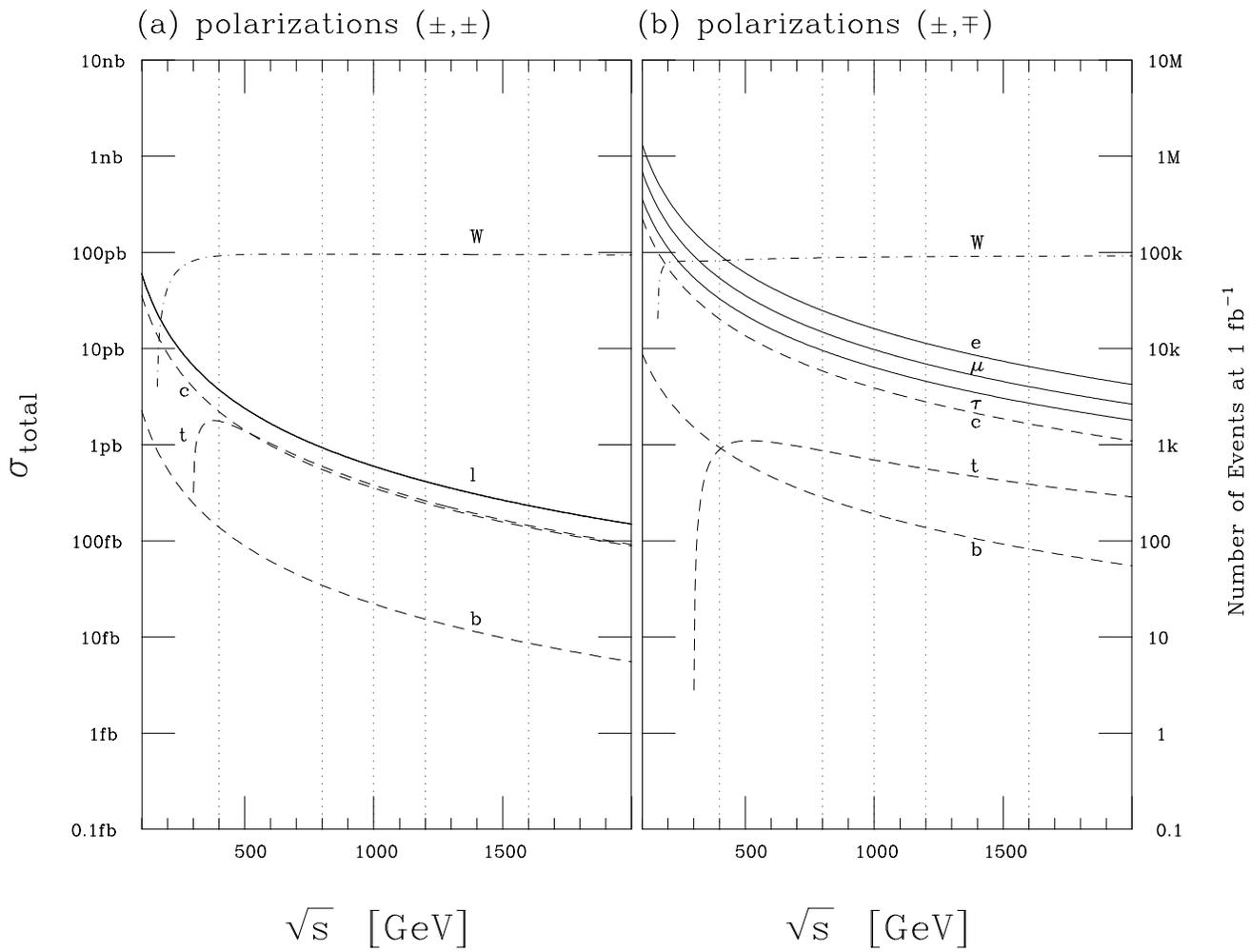
\item[Fig.~4]  The C.~M.~energy dependences of lepton (solid line), quark
     (dashed line) and $W$ boson (dot-dashed line) pair production cross
     sections without the angle cut.  The adopted values of c quark, b quark,
     t quark and $W$ boson masses are 1.5, 5, 150 and 80~GeV, respectively.
     The fine structure constant is fixed to 1/128.  Figures (a) and (b) are
     for the same and the opposite sign photon helicities, respectively.
     All cross sections are given by the tree--level computations.

\newpage

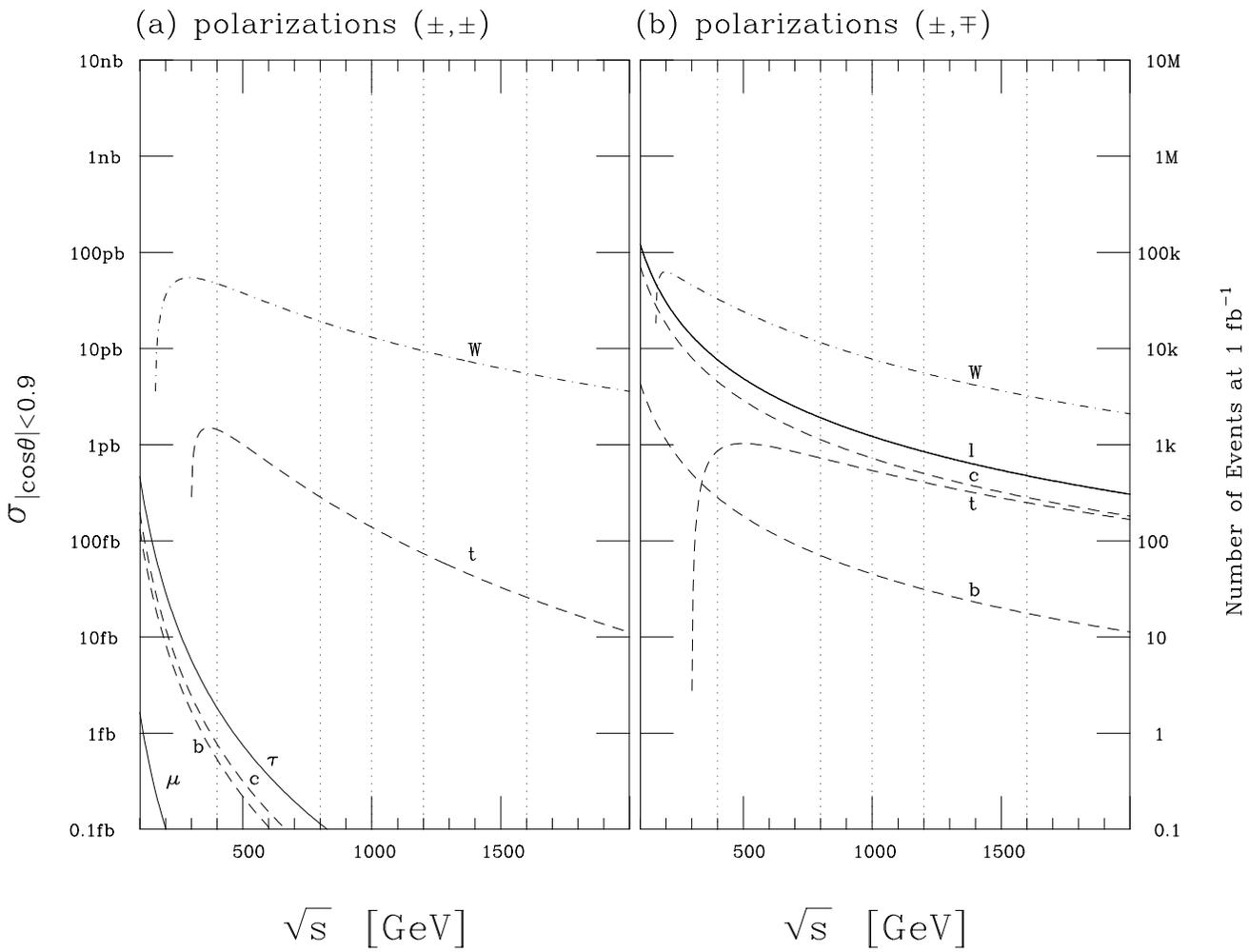
\item[Fig.~5]  The C.~M.~energy dependences of lepton (solid line), quark
     (dashed line) and $W$ boson (dot-dashed line) pair production cross
     sections with an angle cut $| \cos \theta | \leq 0.9$.  The adopted
     values of c quark, b quark, t quark and $W$ boson masses are 1.5, 5,
     150 and 80~GeV, respectively.  The fine structure constant is fixed to
     1/128.  Figures (a) and (b) are for the same and the opposite sign photon
     helicities, respectively.  The line corresponds to electron is far below
     from the figure range in (a).  All cross sections are given by the
     tree--level computations.

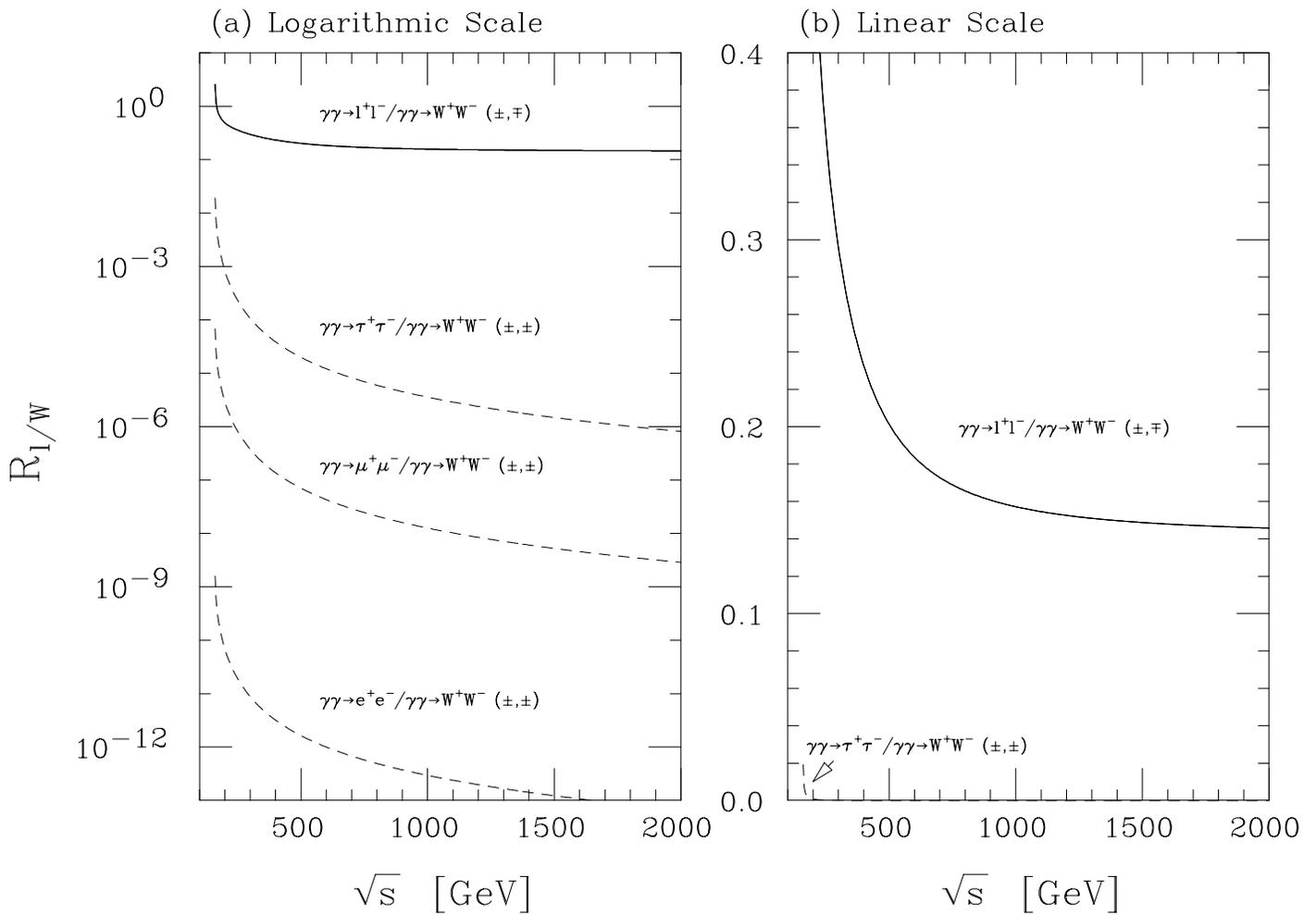
\item[Fig.~6]  The C.~M.~energy dependences of the ratio of pair production
     cross sections of the lepton and $W$ boson.  The dashed (solid) line is
for     the same (opposite) sign photon helicities.  The $W$ boson mass is
     assumed to 80~GeV.  Figure (a) is represented in the logarithmic scale,
     while (b) is in the linear scale.  The angle cut $a$ = 0.9 is adopted.

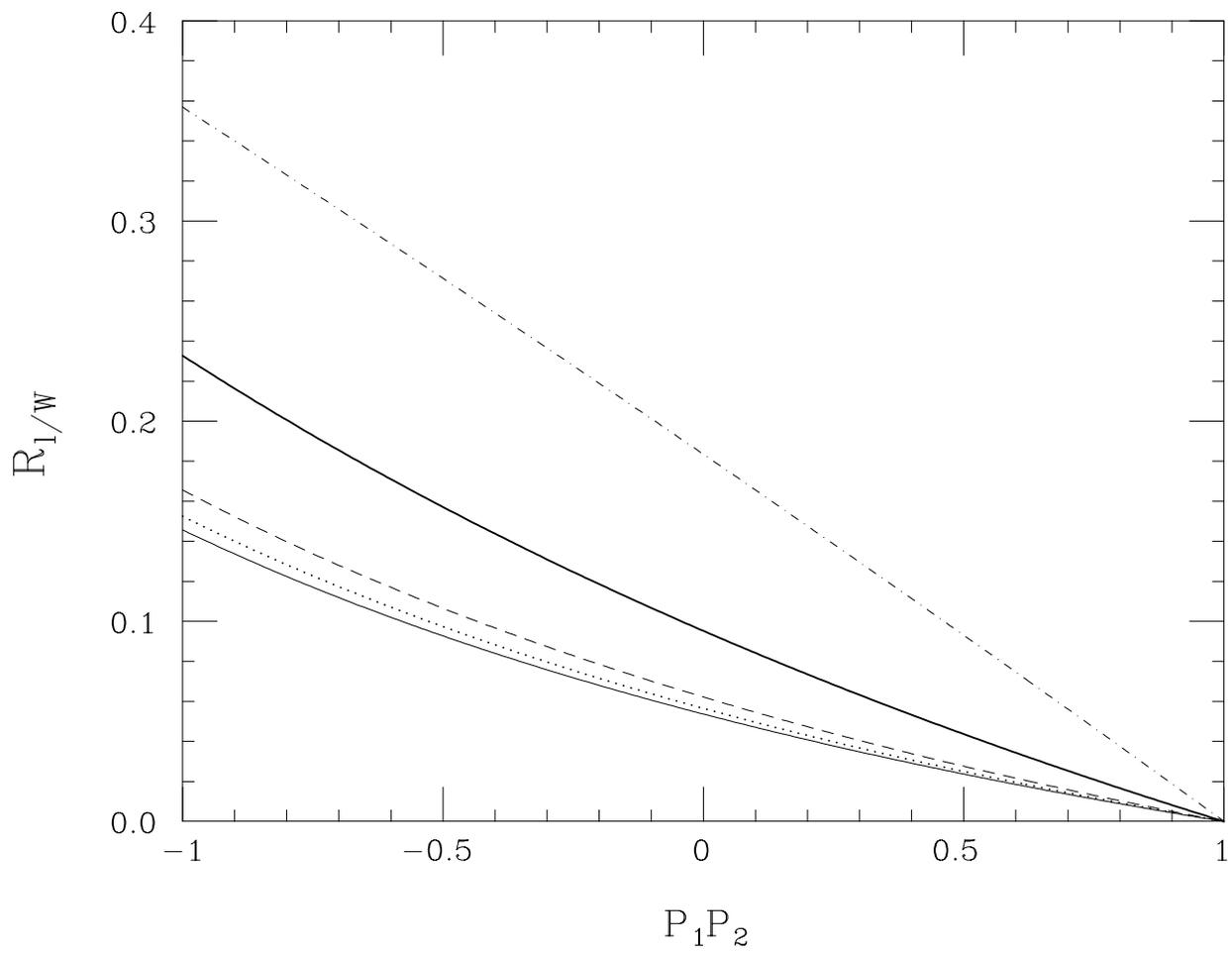
\item[Fig.~7]  The photon beam polarization dependences of the ratio of pair
     production cross section of the light lepton $e$ or $\mu$, to those of
     $W$ boson.  The dot-dashed line corresponds to $\sqrt{s}$ = 250~GeV,
     the bold-solid line 400~GeV, the short dashed line 800~GeV, the dotted
     line 1200~GeV and the slender-solid line 2000~GeV.  The $W$ boson mass is
     assumed to 80~GeV.  The angle cut $a$ = 0.9 is adopted.

\end{description}

\end{document}